\documentclass[letter,longauth]{aa} % for the letters 
\usepackage{graphicx}
\usepackage{xcolor}
\usepackage{float}
\usepackage{chngcntr}
\usepackage{textcomp}
\usepackage{siunitx}
\usepackage{threeparttable}

\usepackage[colorlinks]{hyperref}  
\hypersetup{citecolor=blue} 

\usepackage{txfonts}

\defcitealias{2022aShenar}{TS22}
\def\kms{km\,s$^{-1}$}

\begin{document}

   \title{The Tarantula massive binary monitoring}

   \subtitle{VII. The nature of the eccentric O+BH binary candidate VFTS\,812\thanks{Based on observations collected at the European Southern Observatory under ESO program IDs 112.260M and 114.275G (PI. H. Sana).}}

    \author{K. Deshmukh \inst{\ref{inst:kul},\ref{inst:lgi}} 
          \and 
          H. Sana \inst{\ref{inst:kul},\ref{inst:lgi}}
          \and  
          O. Verhamme \inst{\ref{inst:kul}}  
          \and 
          R. Willcox \inst{\ref{inst:kul},\ref{inst:lgi}}
          \and 
          P. Marchant \inst{\ref{inst:gent}}
          \and 
          T. Shenar \inst{\ref{inst:TelAv}}
          \and 
          F. Backs \inst{\ref{inst:kul}}
          \and \\
          S. Janssens \inst{\ref{inst:tok}}
          \and 
          B. Ludwig \inst{\ref{inst:kul}}
          \and
          L. Mahy \inst{\ref{inst:rob}}
          \and 
          J. O. Sundqvist \inst{\ref{inst:kul}}
          \and
          J. I. Villaseñor \inst{\ref{inst:mpia}}}

   \institute{{Institute of Astronomy, KU Leuven, Celestijnlaan 200D, 3001 Leuven, Belgium \label{inst:kul}}\\ 
              \email{kunalprashant.deshmukh@kuleuven.be}
        \and 
             {Leuven Gravity Institute, KU Leuven, Celestijnenlaan 200D, box 2415
             3001 Leuven, Belgium \label{inst:lgi}}
        \and 
            {Sterrenkundig Observatorium, Universiteit Gent, Krijgslaan 281 S9, B-9000 Gent, Belgium \label{inst:gent}}
        \and 
            {The School of Physics and Astronomy, Tel Aviv University, Tel Aviv 6997801, Israel\label{inst:TelAv}}
        \and 
            {Research Center for the Early Universe, Graduate School of Science, The University of Tokyo, 7-3-1 Hongo, Bunkyo-ku, Tokyo 113-0033, Japan\label{inst:tok}}
        \and 
            {Royal Observatory of Belgium, B-1180 Brussels, Belgium \label{inst:rob}}
        \and
            {Max-Planck-Institut für Astronomie, Königstuhl 17, D-69117 Heidelberg, Germany \label{inst:mpia}}
        }
   \date{Received 6 January 2026 / Accepted 3 February 2026}

  \abstract
   {Massive O-type stars ($M\gtrsim15\,M_\odot$) with an X-ray-quiet black hole (BH) companion represent a crucial stage in the massive binary evolution leading to binary BH mergers. The population of such binaries remains elusive, with $\lesssim5$ candidate or confirmed systems. The Tarantula nebula harbors thousands of massive stars, 2--3\% of which are expected to have BH companions. It is therefore an ideal place to hunt for such systems. We analyzed 30 epochs of VLT/FLAMES IFU high-resolution observations of the H$\delta$ region and archival FLAMES spectroscopy of VFTS\,812, a 17-day single-lined spectroscopic binary with an O4\,V primary and a minimum secondary mass of $5.1\,M_\odot$. Following careful removal of the nebular contamination, spectral disentangling on the new data did not reveal any signature of the hidden companion. We derive $T_\mathrm{eff}=49^{+3}_{-4}$~kK, $\log L=5.7\pm0.1,$ and $v_\mathrm{rot,max}{\rm \,sin\,}i=110^{+25}_{-35}$~\kms\ for the O4~V component, yielding a (single-star) evolutionary mass of  $53^{+6}_{-5}$~M$_\odot$ and an age in the range 0-1.6~Myr. Using injection tests for various luminous artificial companions in our data, we exhaustively ruled out the presence of any luminous signature from a main sequence star more massive than $6\,M_\odot$. We discuss the possible nature of the companion, suggesting that a rejuvenated O star + BH companion is the most suitable scenario to consistently explain the location, (rejuvenated) young age, eccentricity, and lack of companion signature. While this establishes VFTS\,812 as a strong candidate O+BH system, follow-up observations are deemed necessary for a robust confirmation and to search for accretion signatures on the O4~V star.}


   \titlerunning{VFTS\,812: An eccentric O+BH binary candidate}
   
   \maketitle

\section{Introduction}
\label{sec:intro}

Massive binary evolution is a key formation channel for the mergers of neutron stars and black holes (BHs) that are detected via gravitational waves. 
Knowledge of the intermediate OB-type star + BH binary stage would allow us to place key constraints on this evolutionary channel. In the immediate aftermath of the core-collapse of the primary, especially when the OB+BH system has not started to interact, its orbital and dynamical properties retain pristine information \citep{2024VignaGomez,2025Willcox}. 
Around 2--3\% of OB stars are predicted to host compact companions \citep{2020Langer}. However, known X-ray-quiet OB+BH binaries are scarce, with only a handful recently identified \citep[e.g.,][]{2022Mahy,2022Shenar}. While the {\it Gaia} mission is expected to uncover a significant population of OB+BH binaries \citep{2022Janssens,2025Nagarajan}, spectroscopy will remain the main BH hunting tool for systems with short periods or at large distances, such as in the Magellanic Clouds. 

VFTS\,243 is the only confirmed X-ray-quiet OB+BH binary in the Large Magellanic Cloud (\citealt{2022Shenar}).\ It was discovered as part of a much broader effort to characterize a sample of 51 O-type single-lined spectroscopic binaries (SB1) in the Tarantula nebula region \citep[][\citetalias{2022aShenar} hereafter]{2022aShenar}. A significant fraction of that sample was reclassified as double-lined spectroscopic binaries (SB2) based on spectral disentangling, a technique for the separation of component spectra in spectroscopic binaries \citep[e.g.,][]{1995Hadrava}. Others retained their SB1 status or were flagged as uncertain due to nebular contamination impeding the disentangling process. Subsequently, an improved observational approach was deemed necessary to better identify OB+BH systems in the sample.

VFTS\,812, reported as a 17-day eccentric ($e=0.624$) binary with an O4\,V primary weighing about $40\pm7\,M_\odot$ and a minimum secondary mass of $5.1\pm0.7\,M_\odot$, was one such system classified as uncertain (\citetalias{2022aShenar}). A BH companion to the O4\,V primary would likely descend from a similarly or more massive progenitor, and could potentially be massive itself. Given that the minimum secondary mass is in the BH mass regime, we revisited VFTS\,812 with new observations aiming to perform a better disentangling analysis and establish the nature of its companion. In this letter we present the observations, data reduction, disentangling analysis, and possible nature of the unseen companion.

\section{Observations and data reduction}
\label{sec:obs}

Prior data for VFTS\,812 were obtained with the GIRAFFE spectrograph fed with single FLAMES-MEDUSA fibers on the Very Large Telescope (VLT) in the low-resolution modes LR02 and LR03, along with some additional high-resolution data around the H$_\alpha$ region \citep{2011Evans,2017Almeida}. In this work, VFTS\,812 was observed with the integral field unit (IFU) fiber mode of the GIRAFFE spectrograph with the HR03 grating (4032 -- 4203 \si{\angstrom}, $R\approx50 000$). The choice of this setting was motivated by two expected improvements: (i) a better nebular subtraction compared to single-fiber observations and (ii) a higher resolution enabling more robust spectral disentangling. 

Between October 2023 and September 2025, a total of 30 epochs were obtained under ESO programs 112.260M and 114.275G (see Appendix\,\ref{app:0}). Raw data were first corrected for cosmic rays using \texttt{Astro-SCRAPPY}\footnote{\url{github.com/astropy/astroscrappy}} \citep{2001vanDokkum}.
We then used the FLAMES-IFU data reduction recipes within {\it Esorex} version 3.13.7 to reduce all our data. Each IFU comprises 20 individual fibers bundled in a $4\times6$ configuration barring the corners, where each fiber (referred to as a "spaxel" hereafter) records its own spectrum. Under ideal conditions, four spaxels at the center of the IFU should cover the star, and the remaining spaxels can be used to obtain the sky background in its immediate vicinity. Poor seeing or IFU misalignment are two main factors that can adversely affect the data quality. 

Despite measures taken to improve sky subtraction (see Appendix\,\ref{app:1}), it was nontrivial to choose appropriate star and sky spaxels for perfect subtraction. While there are clear improvements compared to the low-resolution single-fiber data, the spatial as well as temporal variability of the nebular line made it challenging to have no residuals after subtraction, and subsequently to obtain spectra appropriate for spectral disentangling. We circumvented this by implementing error spectra in our disentangling approach, which can compensate for imperfect subtraction (see Sect.\,\ref{sec:spedis}). A detailed description of our optimal sky subtraction strategy is discussed in Appendix\,\ref{app:1}. In total, spectra for 26 out of 30 epochs were successfully extracted and used for analysis in this work (Table \ref{tab:obs-summary}.1).

\section{Constraining component masses}

For an SB1 system with a known mass function, an accurate estimate of the primary mass can place strong constraints on the minimum mass of the secondary. VFTS\,812 has a mass function $f=0.0664\pm0.0032\,{\rm M}_\odot$ and a primary spectral type of O4\,V \citepalias{2022aShenar}. The system does not show eclipses in its light curve, making it impossible to determine dynamical masses without an astrometric orbit. 
Previous estimates for VFTS\,812 heavily relied on approximate evolutionary masses; \citetalias{2022aShenar} report $M_1 = 40\pm7\,M_\odot$, which they derived by taking the mean evolutionary mass of all single stars of a similar spectral type from the 30 Doradus sample of \citet{2018Schneider}. 
We revisited the primary mass estimates using detailed atmospheric analysis and subsequent constraints on the secondary mass.

\subsection{Spectroscopic mass of the primary}
\label{sec:specmass}

We compiled all archival and new FLAMES data for VFTS\,812, obtained with the gratings LR02 (397--457 nm), LR03 (450--508 nm), HR03 (403--420 nm), and HR15N (647--679 nm). 
The spectra from each of the gratings were normalized and co-added in the reference frame of the primary star; we clipped out the nebular emission lines before the co-addition. 
We fit {\sc fastwind} atmospheric models \citep{1997Fastwind,2018Sundqvist} to the combined spectrum to obtain stellar parameters such as the effective temperature ($T_{\rm eff}$), surface gravity (log\,$g$), and helium abundance ($Y_{\rm He}$), among others (see the full list in Table \ref{tab:spec_results}). 
The fitting was done with Kiwi-GA, which explores the stellar parameter space, thereby enabling the calculation of statistically derived error margins \citep{brands_r136_2022, brands_x-shooting_2025, verhamme_x-shooting_2024}. 
More details on spectral fitting are discussed in Appendix \ref{app:2}.

Because of a lack of He\,{\sc i} and nitrogen lines, the derived temperature has large error margins. The absence of both N\,{\sc v} and N\,{\sc iv} lines means we can only place an upper limit on the nitrogen abundance. 
The derived spectroscopic mass of the star is  $34_{-7}^{+4}\,M_{\odot}$, slightly lower than the evolutionary mass reported in \citetalias{2022aShenar}.
This is not fully unexpected as spectroscopic masses are often lower than the derived evolutionary masses \citep{herrero_mass_1992}. 
This is potentially due to the missing turbulent pressure in spectral codes such as {\sc fastwind}, which can be significant \citep{2025Moens}. 
Accounting for this turbulence results in an increase in log\,$g$, which can increase the mass significantly \citep[][Verhamme et al. in prep.]{gonzalez-tora_improving_2025}. 
\citet{2025Moens} found a typical velocity corresponding to the turbulent pressure for an O4\,V star to be around 50 $\rm{km\,s}^{-1}$. This velocity does not affect the equivalent width of the lines like the micro-turbulent velocity would and is physically more akin to the macro-turbulent velocity. However, this velocity is not used in the broadening of the line profiles in the formal solution.
If we include this turbulent pressure in {\sc fastwind}, we find an increased spectroscopic mass of $70_{-15}^{+20}\,M_{\odot}$; this stems primarily from the increased log\,$g$ value (the other parameters remain consistent; Table\,\ref{tab:spec_results}).
Since choosing different values will result in significantly different masses, we adopted the parameters derived without including turbulent pressure and considered the evolutionary masses to be the more suitable estimate for VFTS 812.

\subsection{Evolutionary mass of the primary}
\label{sec:evol}

With this new atmospheric analysis, we obtained an effective temperature $T_{\rm eff} = 49.2^{+3.0}_{-3.6}\,{\rm kK}$ and bolometric luminosity log\,$(L/L_\odot) = 5.72^{+0.07}_{-0.09}$. The primary star was not found to have any significant helium or nitrogen enrichment that could indicate past interactions with the companion. We used the BONNSAI Bayesian tool \citep{2014Schneider} to fit single-star stellar tracks from \citet{2011Brott} to the primary using $T_{\rm eff}$ and log\,$(L/L_\odot)$ as inputs. Surface gravity was not included given its strong dependence on turbulent pressure, which is unconstrained. 

BONNSAI successfully replicated all observables consistently within 1$\sigma$, although with a low mass tail of high projected rotational velocity solutions. Given the modest $v\,{\rm sin}\,i$ of $110_{-35}^{+25}$ km\,s$^{-1}$ measured from the atmospheric analysis, we excluded all BONNSAI solutions beyond a rotational velocity of 300 km\,s$^{-1}$. Subsequently, we constrained the evolutionary mass of the O4\,V star to $M_1 = 53.4^{+6.0}_{-5.2}\,{\rm M}_\odot$, the radius $R_1 = 10.1^{+1.4}_{-0.9}\,{\rm R}_\odot$, and estimated an age of $0.89^{+0.71}_{-0.84}$\,Myr (all values are modes). This mass is significantly higher than the corresponding spectroscopic mass ($34_{-7}^{+4}\,M_{\odot}$). Additionally, the evolutionary mass derived using the $T_{\rm eff}-{\rm log}\,(L/L_\odot)$ values with turbulent pressure included is essentially identical ($M_1 = 54.4^{+8.4}_{-5.5}\,{\rm M}_\odot$).

\subsection{Secondary mass and photometric constraints}
\label{sec:m2phot}

We combined the orbital parameters of VFTS\,812 from \citetalias{2022aShenar} with the mass and radius posterior distributions obtained from BONNSAI to obtain a distribution for the minimum secondary mass, obtaining $M_{\rm 2,min} > 5.1 \,M_\odot$ at 5\% significance. Assuming an isotropic orientation of the orbit, we obtained a secondary mass distribution with $M_{\rm 2} > 5.3 \,M_\odot$  at 5\% significance.

VFTS\,812 shows no significant photometric variability or eclipses in OGLE observations (\citetalias{2022aShenar}). The absence of eclipses means we can put an upper limit on the inclination of the binary, assuming that the faint secondary is large enough to eclipse the bright primary. Assuming a main sequence (MS) companion and a mass--radius relation obtained from zero-age MS stars in the BONNSAI catalog, we used the absence of a primary eclipse to further constrain the value of $M_2$ to exceed $5.5\,M_\odot$ at 5\% significance. Photometric constraints are discussed further in Appendix\,\ref{app:4}.

\section{Spectral disentangling}
\label{sec:spedis}

As discussed in Sect.\,\ref{sec:obs}, the individual epoch spectra of VFTS\,812 still have residual features from nebular line subtraction. To minimize the effects of these residuals on the disentangling, we implemented wavelength-dependent error propagation in the shift-and-add technique of \citetalias{2022aShenar}\footnote{\url{github.com/TomerShenar/Disentangling_Shift_And_Add}} (see their Sect. 3.1). We used an error spectrum for each input science spectrum, which makes it possible to assign negligible weight to the region contaminated by the nebular line. For the HR03 setup, the wavelength region from 4103.9 -- 4106.7 \si{\angstrom} was most affected by the nebular line and was thus essentially excluded from the disentangling process. 
A mathematical description of our modification to the shift-and-add technique of \citetalias{2022aShenar} is presented in Appendix\,\ref{app:3}.

\subsection{Searching for a companion}
\label{sec:compsearch}

\begin{figure}
    \centering
    \includegraphics[trim={0.5cm 0cm 2.5cm 1.3cm}, clip,width=0.99\linewidth]{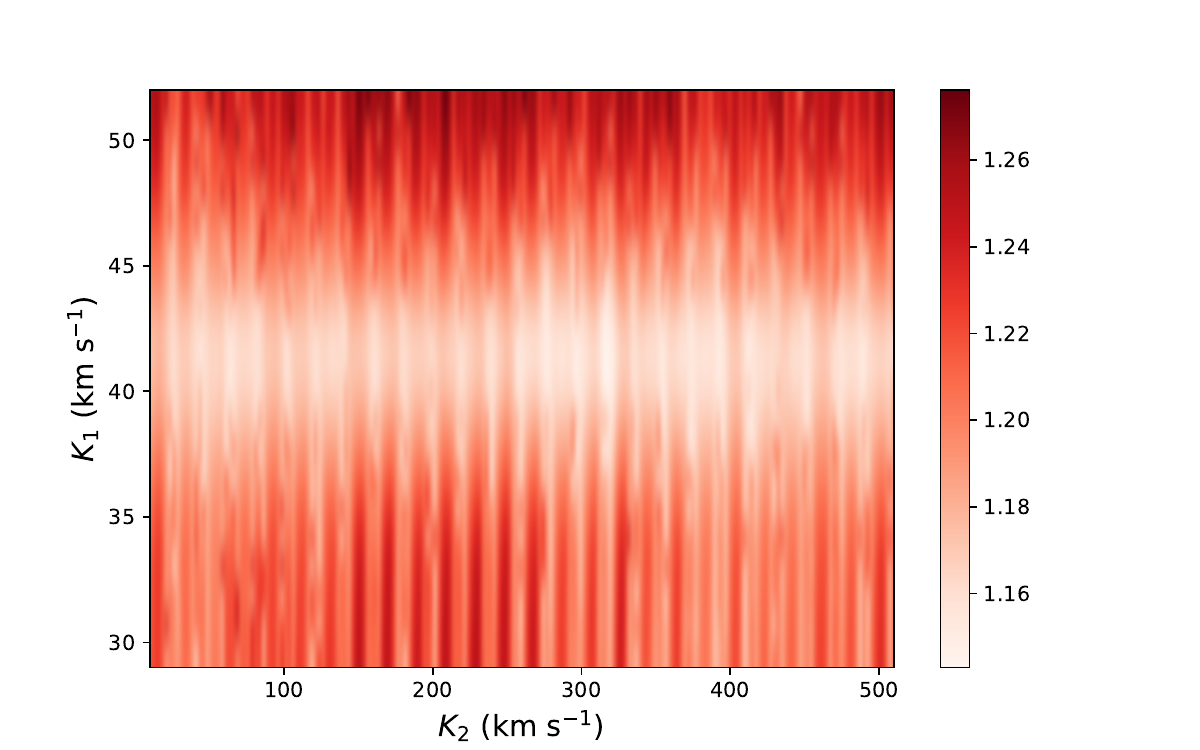}
    \caption{Reduced chi-squared map from grid disentangling as a function of $K_1$ (y-axis) and $K_2$ (x-axis).}
    \label{fig:2D_dis}
\end{figure}

We adopted the orbital solution for VFTS\,812 reported by \citet{2017Almeida}, with an updated time of periastron passage derived from new observations (Table\,\ref{tab:orbelem}). Taking an agnostic approach, we performed two-component disentangling over a grid of component radial velocity (RV) semi-amplitudes, $K_1$  and $K_2$, which range from 30 -- 52 km\,s$^{-1}$ in steps of 1 km\,s$^{-1}$ and 10 -- 500 km\,s$^{-1}$ in steps of 2.5 km\,s$^{-1}$, respectively. This ensured a thorough exploration of different mass ratios. Figure\,\ref{fig:2D_dis} shows a 2D chi-squared map of our disentangling analysis. The primary semi-amplitude, $K_1$, was recovered at $41.9\pm1.0$ km\,s$^{-1}$, consistent with the $42.7\pm0.6$ km\,s$^{-1}$ reported in \citetalias{2022aShenar}. $K_2$, however, was unconstrained, as is evident from the chi-squared map. The disentangled spectrum of the primary is consistent with the O4\,V primary, while that for the secondary is essentially flat and featureless (Fig.\,\ref{fig:1D_dis}). The latter translates to an absence of a detectable companion, either because it is too faint or dark. To further investigate which companion types can avoid detection in our spectra, we carried out an injection study, as described in the following subsection.

\subsection{Limits on the flux and mass of the secondary}
\label{sec:injecstudy}

To test our ability to detect companions of different masses, we performed an injection study wherein we injected companion model spectra into the data and attempted to retrieve them via spectral disentangling. Given the O4\,V primary and assuming the secondary is an MS star, we used BONNSAI to determine the radius, temperature, and luminosity of the injected companion by fixing its age to that of the primary. We computed {\sc fastwind} model spectra based on these inputs, and convolved them with the Bessel B filter to estimate their flux contribution (Table\,\ref{tab:fluxratio}). Subsequently, we synthesized mock spectra by injecting an MS secondary into the data using the computed flux contribution and with an RV semi-amplitude corresponding to its mass.
Mock datasets were made for MS companions of 5, 5.5, 6, 6.5, 7, 8, and 10\,${\rm M}_\odot$. For every dataset, we ran a companion search using disentangling. The disentangled secondary spectrum was retrieved at the same $K_2$ value that it was injected at while keeping the $K_1$ grid the same as presented in Sect.\,\ref{sec:compsearch}. Since the real $K_2$ is not known, we retrieved the secondary spectrum at different $K_2$ values for the no-injection case to search for spectral features.

\begin{figure}
    \centering
    \includegraphics[trim={0cm 0cm 0cm 0.25cm}, clip,width=0.9\linewidth]{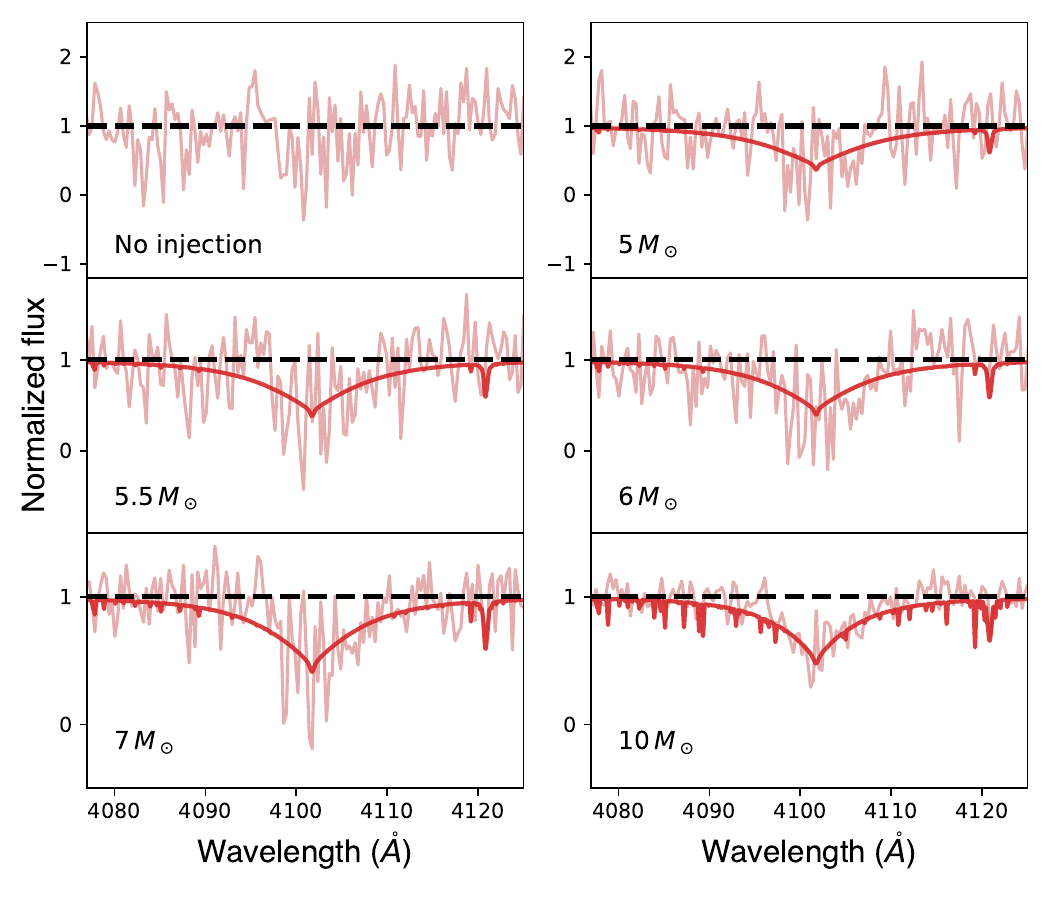}
    \caption{Disentangled secondary spectra (light red) from the no-injection case (first panel) and different mock datasets (remaining panels). Labels give the masses of the injected models. Also shown are the injected model spectra (dark red) and the best-fit flat line (black). The disentangled spectra are binned by a factor of 3 for better visualization.}
    \label{fig:1D_dis}
\end{figure}

The results for select cases are summarized in Fig. \ref{fig:1D_dis}, along with the companion search run from Sect.\,\ref{sec:compsearch} with no injected secondary. Each panel shows the disentangled secondary compared with the injected model (flat for the no-injection case). We performed an $F$-test in the 4072--4130\,\si{\angstrom} range to determine whether the model is a better fit to the retrieved spectrum than a featureless flat line, for which we used a 5\% significance threshold. The continuum region was determined outside the 4090--4112\,\si{\angstrom} region and fixed for both the flat line and model fits. Subsequently, chi-squared values were evaluated in the full 4072--4130\,\si{\angstrom} range to compare the fits.

Assuming rotational velocities of 0, 250, and 500~\kms\ for the injected models, we were able to successfully retrieve companions with masses of 6\,$M_\odot$ and above. For the no-injection case, the flat line versus model fits are statistically indistinguishable when tested with 5, 5.5, and 6 $M_\odot$ models. We can therefore confidently rule out a $\gtrsim6\,M_\odot$ luminous MS companion.

\section{Discussion and conclusion}
\label{sec:discussion}

Based on our disentangling analysis, we discuss here the potential nature of the system and prospects for establishing or rejecting VFTS\,812 as an O+BH binary.
We obtained a detection threshold of $6\,M_\odot$ for MS companions (i.e., lower-mass companions avoid detection). The orbital constraint, on the other hand, implies $M_{\rm 2,min}>5.1\,M_\odot$ at 5\% significance. Subsequently, we considered a few possible scenarios:

MS companion: It is possible that the companion is an MS star with a mass between 5.1 and 6\,$M_\odot$.\ This satisfies the orbital constraint, but any such star remains elusive. Our means for detecting such a companion are limited by current observations. However, in the scenario where VFTS~812 is formed by two young MS stars, its location in 30~Dor remains intriguing: VFTS~812 is offset by  $\sim$4.9\arcmin\ from R136, the star-forming region compatible with the age of VFTS\,812 in the 30~Dor complex \citep{2018Schneider}. VFTS\,812 is also noticeably isolated; the few stars in its neighborhood are much older B-type stars \citep{2018Schneider}. Furthermore, it shows no indication of a runaway motion, neither in its systemic RV \citep{2022Sana} nor in its {\it Gaia} proper motion \citep{2024Stoop}, meaning it could not have been dynamically ejected from R136. 

Hot stripped companion: A hot stripped companion formed via binary mass transfer with $M_2\gtrsim5.1\,M_\odot$ peaks in the ultraviolet (UV) regime and is considerably smaller in radius compared to similar-mass MS stars. It thus contributes even less in optical wavelengths in comparison. An excess flux in the UV is an effective way to detect stripped companions \citep{2025Ludwig}; however, the UV excess contribution of a $\gtrsim5.1\,M_\odot$ stripped star is expected to vanish next to a $53.7\,M_\odot$ MS companion \citep[see Sect. S2.1 in][]{2023Drout}. The photometric AB-magnitude colors of VFTS\,812 using {\it Swift} and Bessel filters (UVM2 -- V = 0.78, UVW1 -- B = 0.64, U -- V = 0.01) are all far from passing the UV excess criteria of \citet{2025Ludwig}. 
For companions with $M_2\gtrsim11\,M_\odot$ where stripped stars transition to the Wolf-Rayet regime \citep{2019Shenar}, the companion would be much more luminous and detectable. On the other hand, for $5.1\,M_\odot \lesssim M_2\lesssim 11\,M_\odot$, factors such as an extreme mass ratio, high eccentricity, no evidence of high rotational velocity, and He or N enrichment in the primary pose significant challenges for this scenario. We therefore consider the presence of a hot stripped companion unlikely.

Compact companion: If the companion is a compact object, the orbital constraint on $M_2$ essentially requires a BH companion. A significant kick during BH formation could explain its isolated location and eccentricity \citep{2022Mahy,2025Willcox}.
Although the primary lacks obvious signatures of a past interaction, if we consider that it was rejuvenated via accretion followed by BH formation with a kick, we can explain its young age, isolated location, and eccentricity.

Fast rotating primary: In Sect.\,\ref{sec:evol} we excluded high rotational velocity ($>300\,{\rm km\,s}^{-1}$) solutions for the primary star from BONNSAI. If the primary is a fast rotator, the following can be said about potential companions: (i) For an MS or stripped companion whose primary rotational axis and orbital axis are aligned, the orbital inclination would be $i\lesssim20^\circ$. This would imply a significantly higher minimum companion mass and a high probability of being detectable. (ii) For a BH companion, a BH formation kick could disturb the above alignment, and similar constraints on its mass could not be made, although the remaining arguments made earlier about a BH companion would still apply.

Given these formation scenario challenges, VFTS\,812 remains a compelling testbed for binary evolution models. The major limitation of our data is the S/N, which holds the key to confirm the presence or absence of a luminous companion. We encourage follow-up high-S/N spectroscopic observations to confirm the nature of the unseen companion in VFTS\,812. Complementary far-UV spectroscopy can also greatly assist in determining accurate atmospheric parameters for the primary.

\bibliographystyle{aa} 
\bibliography{aanda}

\begin{appendix}

\counterwithin{figure}{section}

\section{Acknowledgements}
\begin{acknowledgements}
This research received support from the Flemish Government under the long-term structural Methusalem funding program, project SOUL: Stellar evolution in full glory, grant METH/24/012 at KU Leuven.
SJ acknowledges support from the Japan Society for the Promotion of Science (JSPS) as a postdoctoral fellow. JIV acknowledges support from the European Research Council for the ERC Advanced Grant 101054731.
TS acknowledges support from the Israel Science Foundation (ISF) under grant number 0603225041 and from the European Research Council (ERC) under the European Union's Horizon 2020 research and innovation program (grant agreement 101164755/METAL).

\end{acknowledgements}

\section{Summary of observations}
\label{app:0}

\begin{table}[H]
    \begin{center}
    \caption{FLAMES/IFU observation summary for VFTS\,812.}
    \begin{tabular}{cccc}
         \hline \hline
         Epoch \# & MJD$_{\rm start}$ & Grade & S/N \\
         \hline
         1 & 60219.28064 & C & 28.5 \\
         2 & 60270.16428 & D & 43.2 \\
         3 & 60296.04246 & C & 14.1 \\
         4 & 60313.07963 & C & 41.0 \\
         5 & 60320.15121 & C & 13.6 \\
         6 & 60343.12974 & C & 28.2 \\
         7* & 60356.04970 & C & -- \\
         8 & 60356.05376 & C & 27.1 \\
         9 & 60376.03744 & A & 27.6 \\
         10* & 60379.02918 & A & -- \\
         11 & 60386.02954 & A & 22.5 \\
         12 & 60388.04696 & A & 40.4 \\
         13* & 60390.03915 & A & -- \\
         14 & 60585.31163 & A & 19.4 \\
         15 & 60608.21356 & A & 46.4 \\
         16 & 60644.21282 & A & 25.2 \\
         17 & 60647.21363 & A & 48.1 \\
         18 & 60666.18454 & A & 43.6 \\
         19 & 60674.21820 & A & 49.8 \\
         20 & 60678.05188 & A & 31.0 \\
         21 & 60680.07364 & A & 21.2 \\
         22 & 60683.22568 & A & 23.8 \\
         23 & 60697.17431 & A & 36.6 \\
         24 & 60707.11384 & C & 26.2 \\
         25 & 60709.08383 & A & 16.2 \\
         26* & 60740.10798 & B & -- \\
         27 & 60765.00890 & C & 10.7 \\
         28 & 60768.98720 & B & 20.5 \\
         29 & 60933.35310 & A & 18.6 \\
         30 & 60947.26876 & C & 43.3 \\
         \hline
    \end{tabular}
     \end{center}
    \label{tab:obs-summary}
\tablefoot{For every epoch, we provide the MJD at the start of the observation (Col. 2), grade assigned by ESO (Col. 3), and S/N measured in the continuum region between 4050 -- 4060 \si{\angstrom} (Col. 4). For epochs marked with an asterisk, the data quality was too poor to extract spectra.}
\end{table}

\section{Optimal sky subtraction}
\label{app:1}

\begin{figure}%[H]
    \centering
    \includegraphics[trim={0cm 1cm 0cm 1cm}, clip,width=0.95\linewidth]{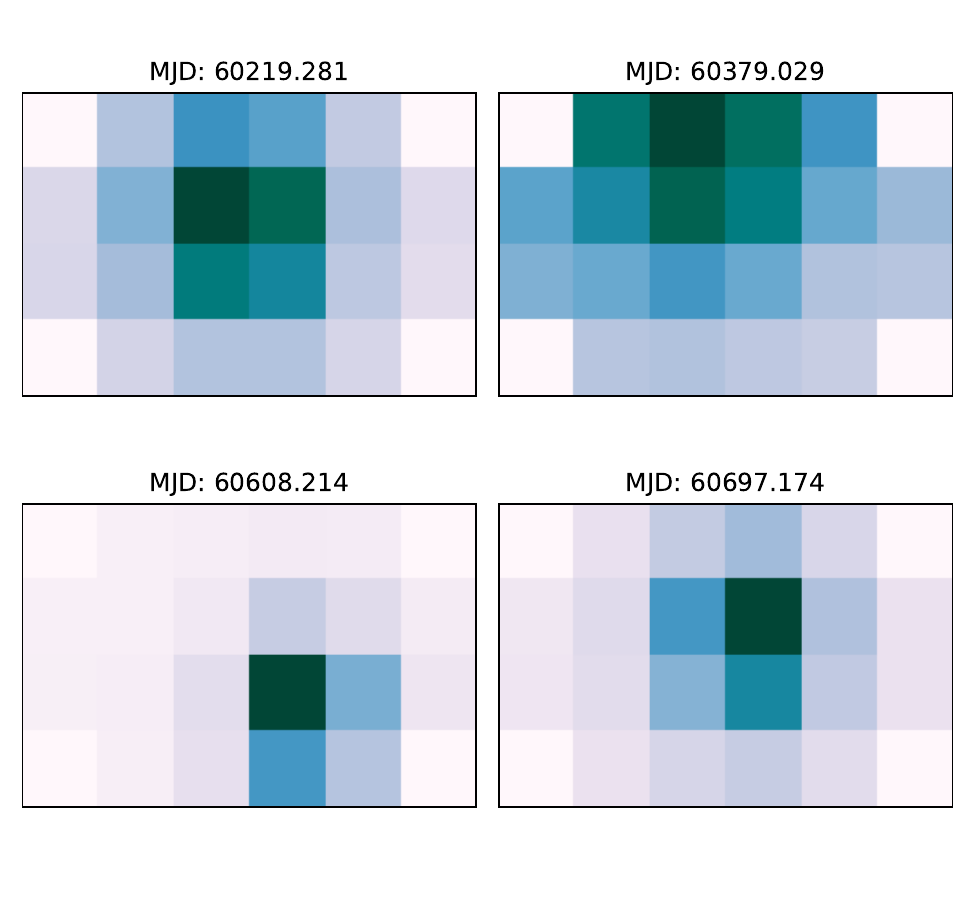}
    \caption{Four examples of IFU data shown with their modified Julian dates (MJDs) at the start of observation. In every panel, the 20 spaxels barring the corners are scaled with their median flux, with light colors representing low and dark representing high flux. The position and point spread function of the star can be seen varying across the four epochs.}
    \label{fig:ifu}
\end{figure}

\begin{figure}
    \centering
    \includegraphics[width=0.95\linewidth]{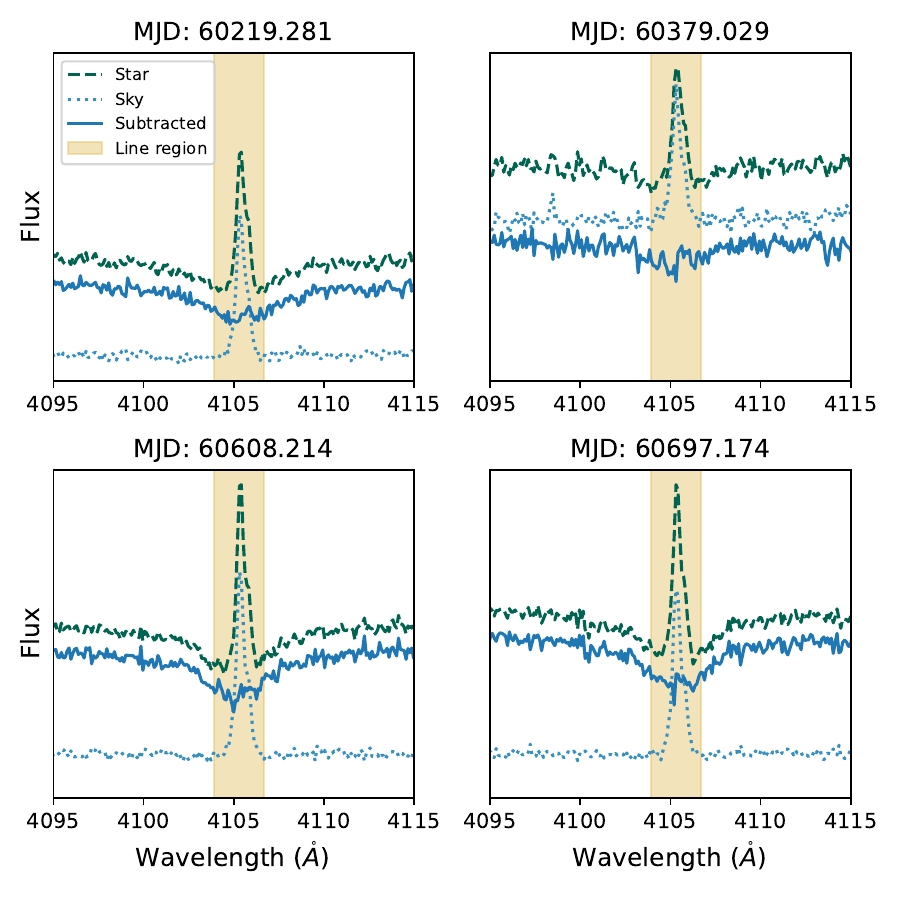}
    \caption{Spectra for the star spaxels (dashed), sky spaxels (dotted), and the resulting sky-subtracted spectrum (solid) for the same four examples as in Fig.\,\ref{fig:ifu}. Also highlighted is the ``line region'' (yellow) from 4103.9--4106.7\,\si{\angstrom} that was most affected by nebular contamination. The sky subtracted spectra in the line region clearly show variable residual features across the four epochs.}
    \label{fig:skysub}
\end{figure}

The spatial and temporal variation in the Tarantula nebula makes it difficult to obtain perfect sky subtraction. The IFU approach is helpful to remedy this by capturing the sky background in the vicinity of the star for every epoch. However, we noticed two important features in our observations: (i) The star is not always aligned on the central four spaxels and changes positions on the IFU from epoch to epoch (Fig.\,\ref{fig:ifu}); and (ii) The nebular emission in the immediate vicinity of the star is not always identical to that in the star itself (Fig.\,\ref{fig:skysub}). Perfect sky subtraction is thus not guaranteed even with IFU observations. This was considered when choosing the spaxels for the star and sky regions. Here, we discuss two approaches along with their suitability for spectral disentangling.

In the first approach, we prioritized perfect nebular subtraction to leverage the capabilities of the IFU mode. For each epoch, we selected the four brightest spaxels for the star and summed them up. To obtain perfect subtraction, we tried different combinations of four other spaxels to sum up and find a suitable sky spectrum. This was done for each epoch individually, leading to varying choices for sky spaxels across epochs. In many cases, we chose sky spaxels immediately next to the star spaxels for the best possible subtraction. This came at the cost of subtracting some of the star signal present in these spaxels, reducing the signal-to-noise ratio (S/N) of the resulting spectrum. Despite these considerations, the nebular subtraction in many spectra remained imperfect and variable across epochs as seen in Fig.\,\ref{fig:skysub}. This can severely hamper spectral disentangling when dealing with faint/unseen companions that require the best quality data to obtain meaningful constraints. Furthermore, the compromise on S/N is yet another factor making disentangling more challenging.

In the second approach, given that the nebular subtraction leaving small residuals was inevitable, we prioritized S/N which is particularly important when searching for a faint companion. We followed the same procedure as before to select the four star spaxels. However, for the sky subtraction, we chose four spaxels farthest from the star that were unlikely to have any significant signal from the star. This resulted in a more accurate sky subtraction for the continuum region, while the nebular subtraction was slightly compromised. To further optimize the process, we took advantage of the fact that the sky background did not have any spectral lines in our wavelength range other than the H$_\delta$ nebular emission line. We divided the sky spectrum into two parts: the H$_\delta$ line spanning the small wavelength range of 4103.9 -- 4106.7 \si{\angstrom}, and the continuum region covering the remaining wavelength range. The latter was modeled with a cubic polynomial and combined with actual data within the H$_\delta$ line to form a new sky spectrum. The main motivation behind doing this was to avoid adding extra continuum noise in the final sky subtracted spectra. Although the nebular subtraction in this case slightly worsened compared to the previous approach, it is circumvented successfully in our modified disentangling method discussed in Sect.\,\ref{sec:spedis}.

\section{Atmospheric analysis}
\label{app:2}

\begin{figure*}
    \centering
    \includegraphics[width=\linewidth]{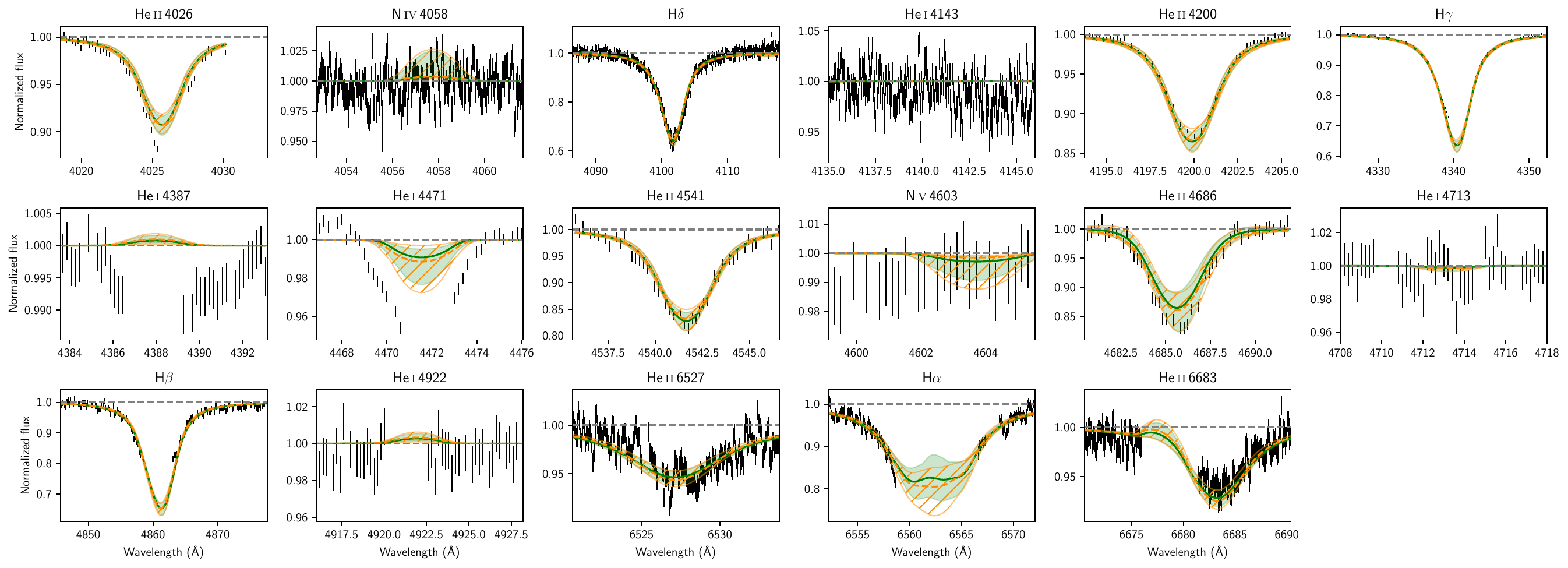}
    \caption{Spectral fit of VFTS\,812. Data is shown in black; the dark green line denotes the best fit, and the light green region the 1$\sigma$ uncertainty region of the fit without turbulent pressure. The dashed orange line and the hatched orange area show the best fit and 1$\sigma$ uncertainty of the fit with turbulence.}
    \label{fig:spec_fit}
\end{figure*}

We use {\sc fastwind} \citep{1997Fastwind,2018Sundqvist} and Kiwi-GA \citep{brands_r136_2022, brands_x-shooting_2025} to fit the spectral lines of VFTS 812.
This method is an evolution from previous genetic algorithm based fitting solutions \citep{mokiem_spectral_2005, tramper_properties_2014, abdul-masih_clues_2019, hawcroft_empirical_2021}.
We allow for 6 fit parameters ($T_{\rm eff}$, $\log_{10}g$, $\dot{M}$, $n_{\rm He}/n_{\rm H}$, $\varv_{\rm rot, max} \sin i$, and nitrogen abundance). 
Here $\varv_{\rm rot, max} \sin i$ is the only broadening mechanism included meaning this value could be read as the combination of macro turbulence and rotational broadening. The micro turbulence is kept fixed at 10 \kms. 
The nitrogen number abundance N is expressed in $\epsilon_{\rm N} = 12 +\log_{10}(n_{\rm N}/ n_{\rm H})$. 
We used {\sc fastwind} version 10.5 \citep{puls_atmospheric_2005} with metallicity scaled to 0.5 times the solar metallicity \citep{asplund_2009} and a smooth wind. 
In total we have access to 17 spectral windows (Fig. \ref{fig:spec_fit}) containing a total of 41 diagnostic spectral lines (Table \ref{tab:linelist}).
In this method all lines were fitted at the same time and the total goodness of fit was taken into account to decide the quality of the fit.
This means that even lines which are generally not thought to be indicative lines of a parameter still contribute to the determination of that parameter. 
To determine the radius of the star, we used the observed K-band 
magnitude of 14.17, an estimated 50 kpc distance to the Large Magellanic Cloud \citep{2019Pietrzynski}, and the $R_V = 3.1$ extinction law \citet{1999Fitzpatrick} to scale the radius such that the model spectral energy distribution crosses the K-band luminosity.
Table \ref{tab:spec_results} lists the best-fit values with their 1$\sigma$ uncertainty.

When using the unaltered version of {\sc fastwind}, turbulent pressure is not included into the calculation of the stellar structure. 
\citet{debnath_2d_2024} showed that the 2D average stellar structure is very different from the 1D models used for spectral synthesis. 
Additionally, they showed that by including turbulent pressure in the hydrostatic equation, it is possible to offset this difference.
\citet{2025Moens} showed through a grid of multi-D models that the turbulent pressure is highly dependent on the Eddington factor of the star, giving us an indication of which turbulent pressure to include. Subsequently, the derived surface gravity, and the spectroscopic mass goes up as the turbulent pressure increases \citep[][Verhamme et al. in prep]{gonzalez-tora_improving_2025}.
All this led us to try a spectral fit in which we included a constant turbulent velocity of 50 \kms. 
The results of this fit did not change any parameters outside of the error-margins 
besides from the surface gravity and thus the resulting derived spectroscopic mass. 
The results of this fit are also given in Table \ref{tab:spec_results} next to the normal fit for ease of comparison. 
Figure \ref{fig:spec_fit} shows the best-fit models with and without turbulent pressure, compared to the data with the 1$\sigma$ range of the spectra.
The spectroscopic mass strongly depends on the chosen turbulent velocity, which is only loosely bound by the radiation hydrodynamical models \citep{2025Moens}.
Because it is not possible to constrain the turbulent velocity observationally, strong constraints on spectroscopic mass are inaccessible for VFTS\,812.  

\begin{table}%{p{2.cm}p{2.cm}p{2.cm}}
        \begin{center}
                \caption{\normalsize{All spectral lines that were fitted. }}\label{tab:linelist}
                \begin{tabular}{c|c|c}
                        \hline \hline
                        Ion& Rest wavelength $[\AA]$&Line window\\
                        \hline
                        He\,{\sc ii} & 4025.4 & He\,{\sc ii} 4026\\
                        He\,{\sc i} & 4026.2 & He\,{\sc ii} 4026\\
            \ion{N}{iv} & 4057.8 & \ion{N}{iv} 4058 \\
                        S\,{\sc iv} & 4088.9, 4116.1 & H$\delta$ \\
                        N\,{\sc iii} & 4097.4, 4103.4 & H$\delta$ \\
                        H\,{\sc i} & 4101.7 & H$\delta$ \\
            \ion{He}{i} & 4143.8 & \ion{He}{i} 4143 \\
                        N\,{\sc iii} & 4195.8, 4200.1, 4215.77 & He\,{\sc ii} 4200\\ 
                        He\,{\sc ii} & 4199.6 & He\,{\sc ii} 4200\\
                        He\,{\sc ii} & 4338.7 & H$\gamma$\\
                        H\,{\sc i} & 4340.5 & H$\gamma$\\
                        O\,{\sc ii} & 4317.1,4319.6, 4366.9 & H$\gamma$\\
                        N\,{\sc iii} & 4345.7, 4332.91 &  H$\gamma$\\
            \ion{He}{i} & 4387.9 & \ion{He}{i} 4387 \\
                        He\,{\sc i} & 4471.5 & He\,{\sc i} 4471\\
                        N\,{\sc iii} & 4534.6 & He\,{\sc ii} 4541\\
                        He\,{\sc ii} & 4541.4 & He\,{\sc ii} 4541\\
            \ion{N}{v} & 4603.7 & \ion{N}{v} 4603 \\
                        He\,{\sc ii} & 4685.6 & He\,{\sc ii} 4686 \\
            \ion{He}{i} & 4713.1 & \ion{He}{i} 4713 \\
                        N\,{\sc iii} & 4858.7, 4859.0, 4861.3,  & H$\beta$\\
                        & 4867.1, 4867.2, 4873.6 &\\
                        He\,{\sc ii} & 4859.1 & H$\beta$ \\
                        H\,{\sc i} & 4861.4 & H$\beta$\\
            \ion{He}{i} & 4921.9 & \ion{He}{i} 4922 \\
            He\,{\sc ii} & 6527.1 & He\,{\sc ii} 6527\\
                        He\,{\sc ii} & 6559.8 & H${\alpha}$\\
                        H\,{\sc i} & 6562.8 & H${\alpha}$\\
                        He\,{\sc i} & 6678.2 & He\,{\sc ii} 6683\\
                        He\,{\sc ii} & 6682.8 & He\,{\sc ii} 6683\\
                        \hline
                        %\normalsize
                \end{tabular}
        \end{center}

\tablefoot{The first column shows the atom and its ionization stage which is responsible for the transition. The second column shows the corresponding rest wavelength with possible multiplets. The third column shows where to find this line in the fit summary (Fig. \ref{fig:spec_fit}).}
\end{table}

\renewcommand{\arraystretch}{1.4}
\begin{table}[]
    \centering
    \caption{Spectral fit values with 1$\sigma$ uncertainty.}
    \begin{tabular}{l|rr}
         \hline \hline
         Parameter                                 & without $\varv_{\rm turb}$  & with $\varv_{\rm turb}$ \\ \hline
         $T_{\rm eff} [{\rm kK}]$                  & $49.2_{-3.6}^{+3.0}$        & $47.6_{-3.0}^{+5.2}$    \\
         $\log_{10} g\,[{\rm cm/s}^2]$             & $3.96_{-0.10}^{+0.08}$      & $4.26_{-0.12}^{+0.16}$  \\
         $\log_{10}(\dot{M}) $[${\rm M}_\odot$/yr] & $-6.50_{-0.15}^{+0.10}$     & $-6.55_{-0.20}^{+0.10}$ \\
         $Y_{\rm He} [n_{\rm He}/n_{\rm H}]$       & $0.10_{-0.03}^{+0.02}$      & $0.09_{-0.02}^{+0.02}$  \\
         $\varv_{\rm rot, max} \sin i$\,[\kms]     & $110_{-35}^{+25}$           & $120_{-35}^{+25}$       \\
         N$[12 + \log_{10}(n_{\rm N}/n_{\rm H})]$  & $5.85_{\downarrow}^{+0.90}$ & $5.80_{\downarrow}^{+1.05}$  \\ \hline
         $\log L$\,[L$_\odot$]                     & $5.72_{-0.09}^{+0.07}$      & $5.68_{-0.08}^{+0.13}$   \\
         $R$\,[R$_\odot$]                          & $10.08_{-0.30}^{+0.39}$     & $10.26_{-0.54}^{+0.34}$ \\
         $M_{\rm spec}\,[{\rm M}_\odot]$           & $34_{-4}^{+5}$              & $70_{-15}^{+20}$        \\
         \hline
    \end{tabular}
    \label{tab:spec_results}
\end{table}

\section{Photometric constraints}
\label{app:4}

\begin{figure}[H]
    \centering
    \includegraphics[trim={0cm 0cm 1cm 1cm}, clip,width=\columnwidth]{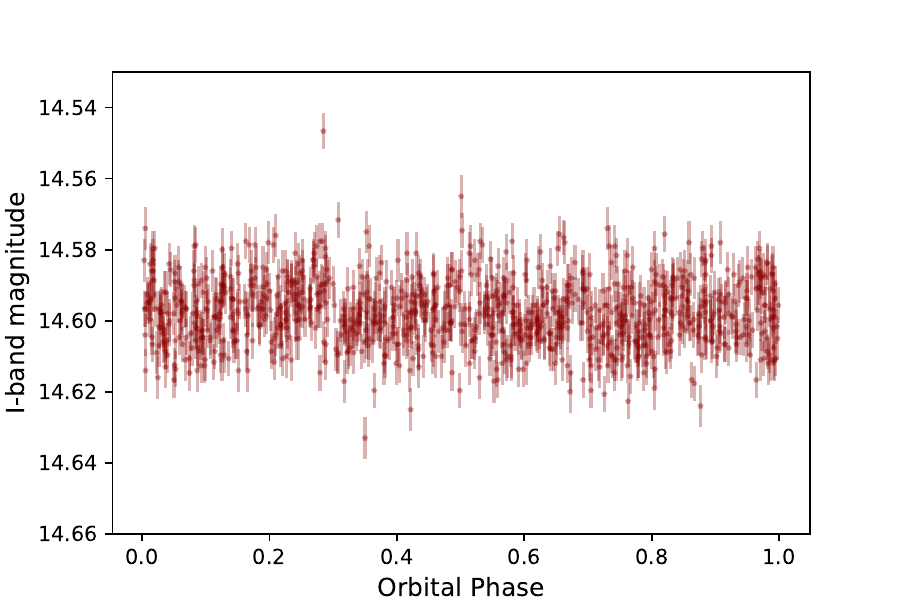}
    \caption{OGLE I-band light curve of VFTS\,812, phase-folded on the orbital period.}
    \label{fig:lc}
\end{figure}

In Sect.\,\ref{sec:m2phot} the lack of observed photometric variability in VFTS 812 was used to obtain an upper limit on the inclination, on the basis that a more edge-on orbit would produce an observable primary eclipse. Figure\,\ref{fig:lc} shows the OGLE I-band light curve that \citetalias{2022aShenar} remarked on for not having any significant frequencies. Using a 3$\sigma$ threshold, we find that the flux-dip from an eclipse has to be $>$\,2.6\% for detection. From spectral disentangling, we know that the flux contribution of the secondary is $\lesssim$\,1.6\% (see Sect.\,\ref{sec:injecstudy}). A full eclipse of the primary by a $5.1\,M_\odot$ companion with a radius of $\sim$$2.3\,R_\odot$ will cover $\sim$5\% of the primary surface, while contributing $\lesssim$\,1.6\% to the total flux, thus resulting in a flux-dip of $\gtrsim$\,3.4\%. While partial eclipses can be considered for finer estimation of eclipse detectability, we adopt the full eclipse scenario to remain conservative. The lack of a detected eclipse therefore enables an additional inclination constraint on the MS companion scenario.

Furthermore, due to the geometry of the system, a stronger inclination constraint could in principle be obtained from the lack of an observed secondary eclipse. However, from spectral disentangling, the brightness of the companion is limited to $\lesssim$1.6\% that of the primary. A secondary eclipse should thus not be observable in OGLE data, and we did not consider this constraint further.

Similarly, the flatness of the light curve also indicates that there is no observed ellipsoidal variability in the primary. Ellipsoidal variability occurs when a star fills a significant fraction of its Roche lobe, thus the absence of such a feature (at fixed primary mass) provides a lower limit on the separation between the components at periastron, and an increase in the minimum secondary mass $M_{\rm 2,min}$ relative to the unconstrained case.
However, for an eccentric orbit, such an analysis requires detailed 3D modeling of the primary and its brightness fluctuations due to regular perturbations from the secondary near periastron. Such modeling was considered out of scope for this work, but may be relevant in future studies.

\section{Spectral disentangling}
\label{app:3}

A comprehensive description of the shift-and-technique is provided in Sect. 3.1 of \citetalias{2022aShenar}. Making an agnostic assumption about the binary being an SB2, a 2D grid disentangling can be used to constrain the respective RV semi-amplitudes $K_1$ and $K_2$ of the primary and secondary components. A chi-squared minimization can yield the best $(K_1,K_2)$ combination, where the chi-squared is calculated as

\begin{equation}
    \chi^2(K_1,K_2) = \frac{1}{N_\lambda(N_{\rm epochs}-2)} \sum^{N_{\rm epochs}}_{i=1} \sum^{N_\lambda}_{k=1} \frac{(A_{i,k}+B_{i,k} - O_{i,k})^2}{\sigma_i^2}.
\end{equation}

Here, $N_{\rm epochs}$ is the number of epochs, $N_\lambda$ is the number of wavelength bins in the selected range of the spectrum, $A_{i,k}$ and $B_{i,k}$ are disentangled spectra and $O_{i,k}$ is the observed spectrum at wavelength $\lambda_k$. The noise term, $\sigma_i$ is calculated in the continuum and used as a representative scalar value. For each combination of $(K_1,K_2)$, the disentangled spectra $A_i$ and $B_i$ are computed iteratively (more details in \citetalias{2022aShenar}).

In case of our IFU observations of VFTS\,812, the imperfect nebular line subtraction led to residuals in the sky-subtracted spectra that can considerably affect the disentangling and introduce artifacts in the disentangled spectra. To mitigate this effect, we change $\sigma_i$ to a vector quantity $\sigma_{i,k}$ defined for all $\lambda_k$. We retain the continuum noise value for most wavelength bins, but arbitrarily increase $\sigma_{i,k}$ to 999 for $\lambda_k \in [4103.9,4106.7]$\,\si{\angstrom}, which is affected by the nebular line as discussed in Appendix\,\ref{app:1}. With this modification, we can perform a meaningful spectral disentangling analysis despite residuals from sky subtraction.

\subsection{Input parameters}

Spectral disentangling relies heavily on an accurate orbital solution; significant deviations from the same can easily introduce artificial features in the disentangled secondary spectrum. The orbital solution for VFTS\,812 was first reported by \citet{2017Almeida} based on 32 epochs of observations covering the wavelength range 3964--4567\,$\AA$ including several spectral lines. They only used He\,{\sc i} and He\,{\sc ii} lines for RV measurement, since Balmer lines are typically broader and affected by nebular emission and wind effects \citep{2013Sana}.
Our new observations in this work are limited to the H$_\delta$ line and are therefore not ideal to find a more constraining orbital solution than \citet{2017Almeida}. Consequently, we adopt their solution in this work, with the only exception being time of periastron passage ($T_0$). With all other orbital elements fixed, we refit $T_0$ using the Spectroscopic and Interferometric Orbital Solution finer (spinOS) tool \citep{2021Fabry} to update the ephemeris and ensure an accurate value for the new data. All orbital elements are summarized in Table\,\ref{tab:orbelem}.

\renewcommand{\arraystretch}{1.2}
\begin{table}[]
    \centering
    \caption{Orbital elements for VFTS\,812.}
    \begin{tabular}{c|c}
         \hline \hline
         Parameter & Value \\
         \hline
         $P$ [days] & 17.28443 $\pm$ 0.00035\\
         $e$ & 0.624 $\pm$ 0.009\\
         $\omega$ [deg] & 339.5 $\pm$ 1.3\\
         $T_{0,{\rm old}}$ [MJD] & 54856.95 $\pm$ 0.04\\
         $T_{0,{\rm new}}$ [MJD] & 60214.87 $\pm$ 0.04\\
         $a\,{\rm sin}\,i$ [$R_\odot$] & 11.39 $\pm$ 0.25\\
         $K_1$ [\kms] & 42.7 $\pm$ 0.6\\
         \hline
    \end{tabular}
    \label{tab:orbelem}
\end{table}

\subsection{Component flux ratios}

For the injection study discussed in Sect.\,\ref{sec:injecstudy}, the mock data need to be synthesized using appropriate flux ratios between the primary and the injected secondary. Similar to \citet{2022Shenar}, we convolve our flux calibrated {\sc fastwind} models with the Bessel B filter and take the ratio subsequently. Table\,\ref{tab:fluxratio} lists the computed flux ratios between different secondary models and the primary model.

\renewcommand{\arraystretch}{1.2}
\begin{table}[]
    \centering
    \caption{Flux ratios for secondary models compared to the primary that were used in the injection study.}
    \begin{tabular}{c|c}
         \hline \hline
         Secondary Mass [$M_\odot$] & Flux ratio [\%] \\
         \hline
         5 & 1.1 \\
         5.5 & 1.3 \\
         6 & 1.6 \\
         6.5 & 1.9 \\
         7 & 2.3 \\
         8 & 3.1 \\
         10 & 5.1 \\
         \hline
    \end{tabular}
    \label{tab:fluxratio}
\end{table}

\end{appendix}

\end{document}